\documentclass[11pt,a4paper]{article} 
\pdfoutput=1
\usepackage{jcappub}

\title{Superconformal generalizations of the Starobinsky model}
\author{Renata Kallosh and Andrei Linde}

\affiliation{Stanford Institute for Theoretical Physics and Department of Physics, \\ Stanford University, Stanford, CA 94305 USA}
 \emailAdd{kallosh@stanford.edu}\emailAdd{alinde@stanford.edu} 
\abstract{We  find a way to represent the Starobinsky model in terms of a simple conformally invariant theory with spontaneous symmetry breaking. We also present a superconformal theory, which, upon spontaneous breaking of the superconformal symmetry, provides a consistent supergravity generalization of  the Starobinsky model.
}

\newcommand{\rf}[1]{(\ref{#1})}

\def\be{\begin{equation}}
\def\ee{\end{equation}}
\def\ba{\begin{eqnarray}}
\def\ea{\end{eqnarray}}
\def\K{K{\"a}hler}

\usepackage{amssymb,graphicx,subfigure}

\begin{document}
\maketitle

\section{Introduction}

Recent investigations in inflationary cosmology revealed a set of rather unexpected facts which require some kind of explanation: Inflationary potentials of several apparently unrelated inflationary models have been shown to coincide, either exactly, or at least during inflation. Therefore (up to some minor corrections related to different mechanisms of reheating) these models lead to identical predictions for the parameters $n_{s}$ and $r$, which nicely fit observational data \cite{Ade:2013rta}.

In particular, a modified version \cite{Starobinsky:1983zz} of the original Starobinsky model \cite{Starobinsky:1980te} has the Lagrangian 
\be\label{star}
L={ \sqrt{-g}} \left({1\over 2} R+{R^2\over 12M^2}\right) \, ,
\ee
where $M \ll M_{p}$ is some mass scale; we will keep $M_{p}= 1$ in what follows.
This theory is conformally equivalent to canonical gravity plus a scalar field $\phi$ \cite{Whitt:1984pd}.
Making the transformation $\tilde{g}_{\mu\nu} = (1 + \phi/3M^2) g_{\mu\nu}$
and the field redefinition $\varphi = \sqrt{\frac{3}{2}} \ln \left( 1+ \frac{\phi}{3 M^2} \right)$,
one finds the equivalent Lagrangian 
\be\label{whitt}
L=\sqrt{-\tilde{g}}\left[{1\over 2}\tilde{R} - {1\over 2}\partial_{\mu} \varphi\partial^{\mu} \varphi - \frac{3}{4} M^2 \left(1- e^{-\sqrt{2/3}\,\varphi}\right)^2 \right] \, .
\ee
Interestingly, the same potential emerges in several other seemingly unrelated models.  It appears, for example, in the Higgs model with a non-minimal coupling to gravity, ${\xi\over 2} \phi^2  R -{\lambda\over 4}(\phi^2-v^{2})^{2}$, for $\xi <0$, in the limit $1 + \xi v^{2} \ll  1$ \cite{Linde:2011nh}.  The potential of the Higgs inflation in the theory ${\lambda\over 4}\phi^4$ with a sufficiently large non-minimal coupling to gravity ${\xi\over 2}\phi^{2}R$ \cite{Salopek:1988qh,Bezrukov:2008ut,Okada:2010jf,Einhorn:2009bh,Ferrara:2010yw,Lee:2010hj,Ferrara:2010in} has the same shape  during inflation. Recently it was found that the same potential  appears in no-scale supergravity, under certain assumptions about moduli stabilization \cite{Ellis:2013xoa}.

As a result, we have a family of  models which may seem totally different but lead to identical or nearly identical observational predictions. To examine possible relations between all of these models, we will study Starobinsky model and its possible generalizations. To our  surprise, we found that this model can be represented as a result of spontaneous symmetry breaking in a very simple conformally invariant scalar field theory.

We will also propose several different ways to embed Starobinsky model to superconformal theory and supergravity. All of them lead to the inflaton potential as in (\ref{whitt}) but only some of them can be consistently stabilized along the inflationary trajectory. As we will see, one of the simplest ways to do it is related to the general approach to chaotic inflation in supergravity proposed in \cite{Kallosh:2010ug}.

 \section{de Sitter from  spontaneously broken conformal symmetry}\label{dSsection}

As a first step towards the new formulation of the Starobinsky model, consider a simple  conformally invariant model of gravity and two real scalar fields, $\phi$ and $h$, which has an $SO(1,1)$ symmetry:
\begin{equation}
\mathcal{L}_{\rm toy} = \sqrt{-{g}}\left[{1\over 2}\partial_{\mu}\chi \partial^{\mu}\chi  +{ \chi^2\over 12}  R({g})- {1\over 2}\partial_{\mu} \phi\partial^{\mu} \phi   -{\phi^2\over 12}  R({g}) -{\lambda\over 4}(\phi^2-\chi^2)^{2}\right]\,.
\label{toy}
\end{equation}
The field $\chi(x)$ is referred to as  a conformal compensator, which we will call `conformon.' This
theory is locally conformal invariant under the following
transformations: 
\be g'_{\mu\nu} = e^{-2\sigma(x)} g_{\mu\nu}\,
,\qquad \chi' =  e^{\sigma(x)} \chi\, ,\qquad \phi' =  e^{\sigma(x)}
\phi\ . \label{conf}\ee 
The field $\chi$ (with negative sigh kinetic term)
can be removed from the theory by fixing the gauge symmetry
(\ref{conf}), for example by taking a gauge $\chi =\sqrt 6$. This gauge fixing can be interpreted as a spontaneous breaking of conformal invariance due to existence of a classical field $\chi =\sqrt 6$.

For our purposes, however, it is much more convenient to choose an  $SO(1,1)$ invariant conformal gauge
\be
\chi^2-\phi^2=6 \ ,
\ee
which reflects the $SO(1,1)$ invariance of our model. This gauge condition represents a hyperbola which can be parametrized by a canonically normalized field $\varphi$, 
\be
\chi=\sqrt 6 \cosh  {\varphi\over \sqrt 6}\, , \qquad \phi= \sqrt 6 \sinh {\varphi\over \sqrt 6} \ .
\ee
In this gauge the Higgs-type potential ${\lambda\over 4}(\phi^2-\chi^2)^{2}$ turns out to be a cosmological constant $9\lambda$,
and our action \rf{toy} becomes
\begin{equation}\label{LE}
L = \sqrt{-g} \left[  \frac{1}{2}R - \frac{1}{2}\partial_\mu \varphi \partial^{{\mu}} \varphi -   9 \lambda  \right].
\end{equation}
This theory has the constant potential $V = 9\lambda$. Therefore it can describe de Sitter expansion with the Hubble constant $H^{2} = 3\lambda$.


\section{Starobinsky  model as a theory with spontaneously broken conformal symmetry}\label{starsection}

Now that we mastered the art of working with  models of this type, we will consider a model which is only slightly different from the dS model described above:
\begin{equation}
\mathcal{L}_{\rm S} = \sqrt{-{g}}\left[{1\over 2}\partial_{\mu}\chi \partial^{\mu}\chi  +{ \chi^2\over 12}  R({g})- {1\over 2}\partial_{\mu} \phi\partial^{\mu} \phi   -{\phi^2\over 12}  R({g}) -{\lambda\over 4}\phi^{2}(\phi-\chi)^{2}\right]\,.
\label{toystar}
\end{equation}
The only difference here is that we replaced the $SO(1,1)$ symmetric potential ${\lambda\over 4}(\phi^2-\chi^2)^{2}$ by the potential ${\lambda\over 4}\phi^{2}(\phi-\chi)^{2}$. This theory has the same conformal symmetry (\ref{conf}), but the new potential breaks the $SO(1,1)$ symmetry of the previous theory.

We will use the same gauge $\chi^2-\phi^2=6$ as in the previous section, with the same relation between the fields $\chi$, $\phi$ and the canonically normalized field $\varphi$:
$\chi=\sqrt 6 \cosh  {\varphi\over \sqrt 6}$, $\phi= \sqrt 6 \sinh {\varphi\over \sqrt 6}$. By using these relations, one can easily show that our new theory has the action
\begin{equation}\label{Starmodel}
L = \sqrt{-g} \left[  \frac{1}{2}R - \frac{1}{2}\partial_\mu \varphi \partial^{\mu} \varphi -   \frac{9\lambda}{4} \left(1- e^{-\sqrt{2/3}\, \varphi}\right)^2 \right].
\end{equation}
It  coincides with the Starobinsky model (\ref{whitt}) after identification $M^{2} = 3\lambda$. In other words, one can represent the Starobinsky model as a simple conformally invariant theory (\ref{toystar}) with spontaneous symmetry breaking.

\section{Supersymmetry and Starobinsky model}

As a next step, we will investigate a possibility to embed Starobinsky model into superconformal theory and supergravity.
Earlier attempts to do it go back to the paper by Cecotti \cite{Cecotti:1987sa}. The author proposed a corresponding superconformal theory, generalizing the bosonic action \rf{star}. However, the resulting supergravity model required introduction of two complex scalar fields. One can show that inflationary regime in this model is unstable with respect to a tachyonic instability of one of these fields, see a discussion in the end of this section. 
Subsequent attempts were made by Ketov et al in \cite{Ketov:2010qz}, where supersymmetrization involved additional terms of higher order in $R$. The relevant supergravity scalar potential depends on a complex `scalaron'.  Its stability during inflation was not investigated, which makes this approach incomplete.
Our present investigation was stimulated by the recent paper \cite{Ellis:2013xoa}, where it was argued that under certain assumptions about supergravity/string theory moduli stabilization, the Starobinsky model  (\ref{whitt}) can be derived from the no-scale supergravity with the Wess-Zumino superpotential. It would be nice to have an explicit realization of this scenario including a specific mechanism of moduli stabilization.\footnote{We are grateful to Keith Olive for a discussion of this issue.}

In this paper we will provide a consistent implementation of the Starobinsky inflationary model in superconformal theory and supergravity using methods developed in \cite{Kallosh:2000ve}  and in \cite{Ferrara:2010yw,Ferrara:2010in,Kallosh:2010ug,delta}. The general approach to supergravity is based on the superconformal theory, which becomes supergravity after spontaneous symmetry breaking, for example,  when the complex conformon field $X^{0}$ acquires a constant value $X^{0} = \sqrt 3$.  This is very similar to the mechanism discussed in the previous section. Here we will only briefly remind the basic features of this approach and present the new results related to the Starobinsky model.

In the simple theory of real scalar fields described in the previous section, the requirement of conformal symmetry resulted in a unique coupling of the inflaton field to gravity, $-{\phi^2\over 12}  R$. In superconformal theory one has many other possibilities to reach a similar goal.  We will use the theory which has three chiral multiplets $X^I$: the compensator field $X^0$, the inflaton $X^1=\Phi $ and the Goldstino  superfield $X^2=S$. 
This theory is characterized by the potential of the embedding manifold $\mathcal{N}(X,\bar X)$ and the superpotential ${\cal W}(S, X^0, \Phi)$, which are related to the more familiar \K\ potential and the superpotential of $N = 1$ supergravity as follows:
\be
\mathcal{N}(X,\bar X)_{X^0=\bar X^{\bar 0}= \sqrt 3}= - 3\,  e^{-{1\over 3}  K(\Phi, \bar \Phi; S, \bar S)} ,\qquad {\cal W}(S, X^0, \Phi)_{X^0= \sqrt 3}  = W(S,\Phi) \ .
\ee
One can embed the Starobinsky model into the superconformal theory in several different ways, but not all of them lead to a theory which is stable during inflation. We will consider a special class of the functions $\mathcal{N}(X,\bar X)$, ${\cal W}(S, X^0, \Phi)$:
\be
\mathcal{N}(X,\bar X)= - \left | X^{0}\right| ^{2}\,  \exp \left(- {\left |S\right| ^{2}\over \left | X^{0}\right| ^{2}} + {1\over 2}  \Bigl({\Phi\over X^0} - {\bar \Phi\over \bar X^{\bar 0}} \Bigr)^2+\zeta{\left |S\right| ^{4}\over \left | X^{0}\right| ^{4}}\right)\  ,
\ee
 \be
 {\cal W}(X^0, \Phi, S)= {M\over 2\sqrt 3} \, S  \, (X^{0})^{2}\, \left(1-e^{-2\Phi/X_{0}}\right) \ .
\label{superpotCal}
\ee
The corresponding \K\, potential and superpotential in supergravity are  \cite{Kallosh:2010ug}
\be
K=  S\bar S- {(\Phi-\bar \Phi)^2\over 2} -\zeta (S\bar S)^2, \qquad W =   {M\sqrt 3\over 2} S \Big (1- e^{- {2 \Phi \over \sqrt 3}} \Big )\, .
\label{Kquadratic} \ee

This theory belongs to the broad class of chaotic inflation models in supergravity developed in  \cite{Kallosh:2010ug}, with the \K\ potential and superpotential  of the general form $K((\Phi-\bar\Phi)^2,S\bar S)$, ${W}= Sf(\Phi)$. Here $f(\Phi)$ can be any real holomorphic function such that $\bar f(\bar \Phi) = f(\Phi)$. Any function which can be represented by Taylor series with real coefficients has this property.  

This class of theories has a number of useful features. The role of the inflaton field in this model is played by the real part $\varphi$ of the field $\Phi = (\varphi+i\alpha)/\sqrt 2$. The \K\ potential is shift-symmetric with respect to $\varphi$. The scalar potential is invariant under rotation $S\to S e^{i\theta}$ and sign change $\alpha \to -\alpha$. Therefore for any given $\varphi$, the potential has an extremum at $S, \alpha  = 0$. If one can achieve stability of $S$, and $\alpha$ near $S, \alpha  = 0$ by making this extremum a minimum (which can be done by a proper choice of the \K\ potential  \cite{Kallosh:2010ug}), the inflationary trajectory will correspond to evolving $\varphi$, with $S, \alpha  = 0$. All fields in this class of models are canonically normalized along the inflationary trajectory. The inflaton potential in this theory is $V=|f(\Phi)|^2$. Therefore the inflaton potential in the model (\ref{Kquadratic}) is given by
\be \label{Starlambda}
V = {3M^{2}\over 4} \Big (1- e^{- \sqrt{2\over 3}\varphi} \Big)^{2} \ .
\ee
Thus the Lagrangian of the inflaton field $\varphi$ in this model exactly coincides with the Lagrangian of Starobinsky model (1.1), (1.2), but  now this model is a part of supergravity and the underlying superconformal theory. It remains to check whether all other fields, such as $S$ and $\alpha$, vanish during the cosmological evolution.

Following  \cite{Kallosh:2010ug}, one can show that the field $\alpha$ is firmly stabilized at $S = 0, \alpha = 0$, with the mass squared of this field asymptotically approaching $6H^{2}$ during inflation at large $\varphi$.  The field $S$ is also stable near $S = 0$, with the mass squared of this field asymptotically approaching $12\zeta H^{2}$. 

\begin{figure}[h!]
\centering
\subfigure[]{
\includegraphics[scale=.29]{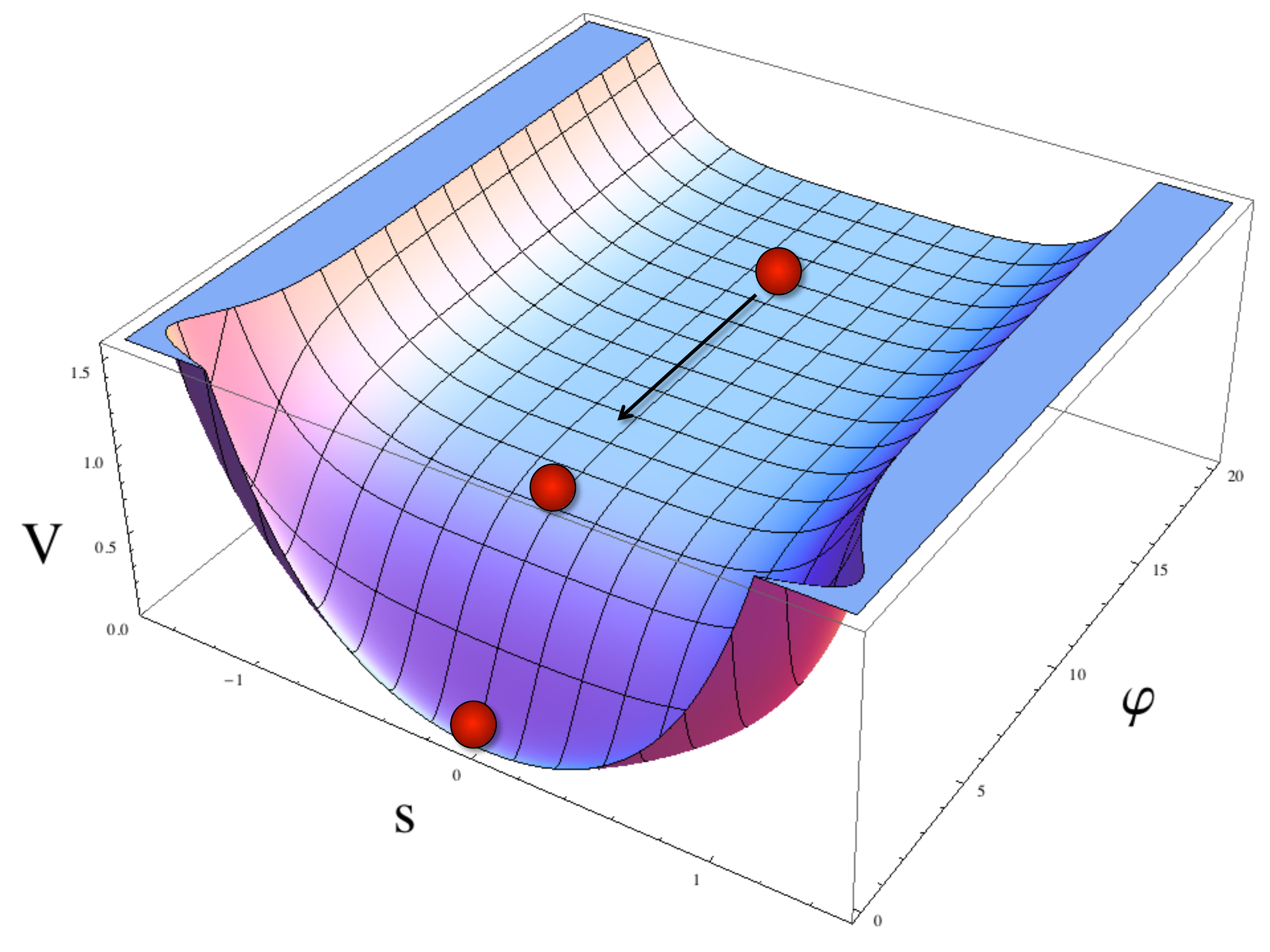}
\label{Star1}}
\subfigure[]{
\includegraphics[scale=.29]{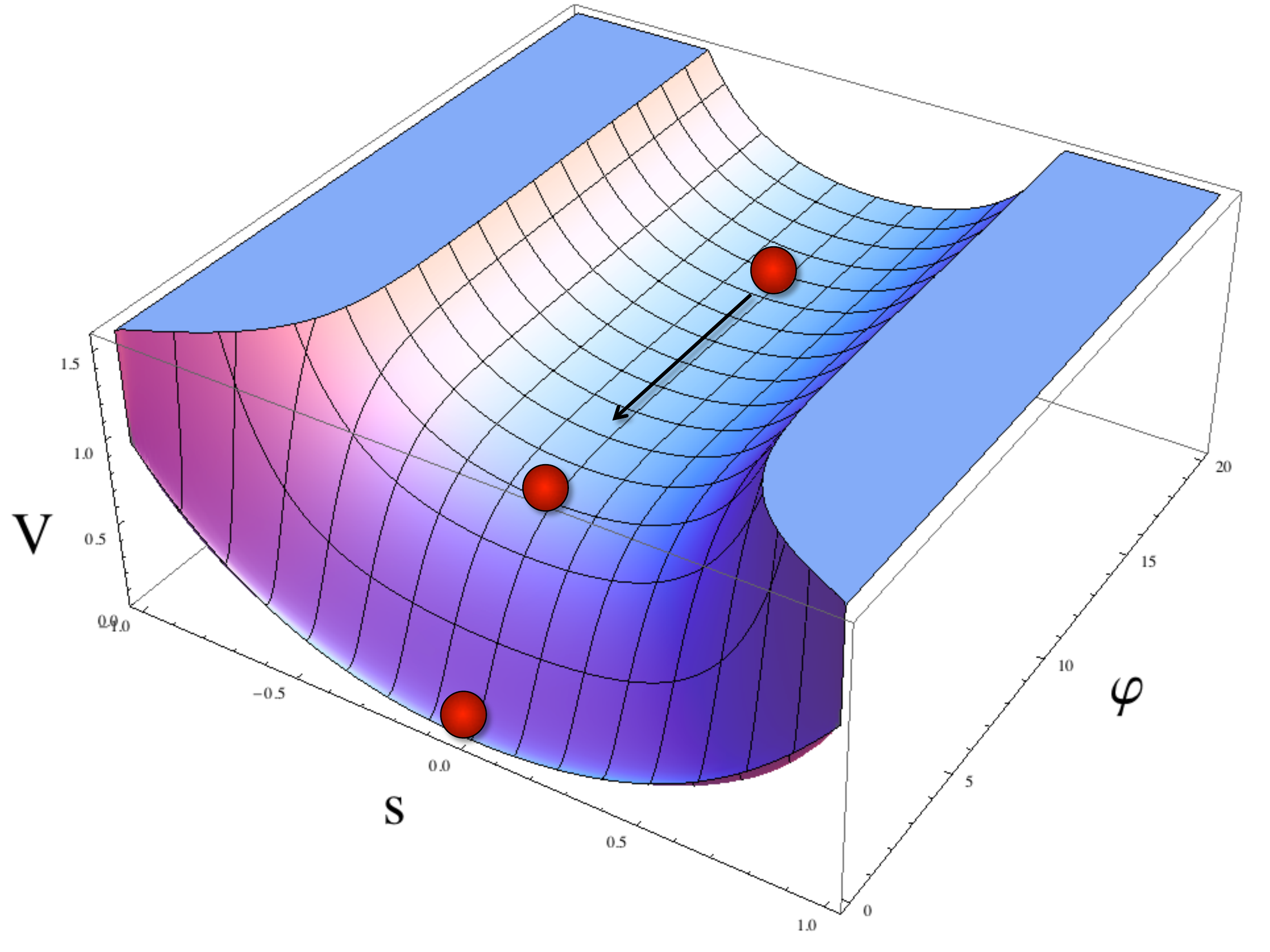}
\label{Star2}}
\caption{The potential of the stable supergravity generalization of the Starobinsky model with the \K\ potential (\ref{Kquadratic}) for $\zeta = 0$, Fig. \ref{Star1}, and for  $\zeta = 0.5$, Fig. \ref{Star2}.  The fields are shown in Planckian units, the height of the potential is shown in units of $\lambda = 1/9$.}
\label{Star}
\end{figure}

Note that the field $S$ is not tachyonic and inflationary regime is described by the Starobinsky model even in the absence of the stabilizing term $\zeta (S\bar S)^2$, see Fig. \ref{Star2}. However, in this regime $m^{2}_{s} \ll H^{2}$, and therefore quantum fluctuations of the field $S$ are generated during inflation, which can be used e.g. for generation of small non-gaussian perturbations via the curvaton mechanism \cite{Demozzi:2010aj}. The field $S$ is firmly stabilized and fluctuations of this field are not generated during inflation if $12\zeta H^{2} \gtrsim H^{2}$, i.e. for $\zeta \gtrsim 1/12$, see Fig. \ref{Star2}.

We should emphasize that this embedding of the Starobinsky model to supergravity is not unique. In particular, one may achieve a similar goal in a theory with
\be
K = -3\log \Bigl[ 1 + \frac{1}{6}(\Phi -\bar\Phi)^2 - \frac{S \bar S}{3} +{\zeta\over 3} (S\bar S)^2  \Bigr] , \quad W =   {M\sqrt 3\over 2} S \Big (1- e^{- {2 \Phi \over \sqrt 3}} \Big )\, .
\label{Kquadratic2} \ee

One may also implement the Starobinsky model in a somewhat modified  Cecotti model \cite{Cecotti:1987sa}, which can be made stable during inflation. An improved expression of the \K\ potential and superpotential for the supergravity version of this model is
\be
K=  -3 \log \left(\Phi+\bar \Phi -S\bar S+{\zeta\over 3} (S\bar S)^2\right), \qquad W =   {3M}\, S\, (\Phi-1)\, .
\label{Cecotti} \ee
Just as in the models (\ref{Kquadratic}), (\ref{Kquadratic2}), one can show that the inflaton direction in this model corresponds to the real part of the field $\Phi$, and the potential of the canonically normalized field $\varphi$ has the Starobinsky potential (1.2). In the original version of this model, the field $S$ was tachyonic during inflation, with the ``conformal'' mass squared $m_{s}^{2} = -2H^{2} = -R/6$. As a result, the inflationary trajectory at $S= 0$ was unstable. Fortunately we found that this problem can be solved by adding the stabilization term $\zeta (S\bar S)^2$, if one takes $\zeta > 0.15$. In order to have $m_{s}^{2} \gtrsim H^{2}$, one should take $\zeta \gtrsim 0.5$. 

\section{Discussion}

In this paper we continued our previous work of implementing inflationary models favored by the Planck2013 data \cite{Ade:2013rta} in the context of advanced theories based on supersymmetry and conformal invariance. In our previous paper we considered various versions of the theory $\lambda\phi^{4}$ with nonminimal coupling to gravity \cite{delta}, and explained the mechanism of implementation of this theory in superconformal theory and supergravity. In this paper we studied the same issue with respect to the Starobinsky model. We found a way to represent the Starobinsky model as a simple locally conformally invariant theory with spontaneous symmetry breaking  (\ref{toystar}). 

We also incorporated the Starobinsky model in the context of superconformal theory and supergravity. The main lesson here is that one can do it in many different ways. However, not all of the supergravity generalizations of the Starobinsky model, as well as of other, more general models of modified gravity $f(R)$  lead to a consistent cosmological theory. The corresponding investigation must include a full analysis of stability of inflationary regime with respect to other fields which appear as a result of a supersymmetric generalization of the original theory. This investigation can be quite involved, but fortunately there are some classes of supergravity models where it can be done in a relatively straightforward way. In this paper we used the general approach to this issue developed in \cite{Kallosh:2010ug}, which allows to confirm the stability of the inflationary trajectory in various cosmological models, including the supersymmetric generalization of the Starobinsky model (\ref{Kquadratic}), using relatively simple analytical methods.

\begin{acknowledgments}
We are very grateful to K. Olive and A. Van Proeyen for many interesting discussions.
This work  is supported by the SITP and by the 
NSF Grant No. 0756174. The work of RK is also supported by the Templeton Foundation Grant ``Frontiers of Quantum Gravity''.  
\end{acknowledgments}


\begin{thebibliography}{99}

\bibitem{Ade:2013rta} 
  P.~A.~R.~Ade {\it et al.}  [ Planck Collaboration],
``Planck 2013 results. XXII. Constraints on inflation,''
  arXiv:1303.5082 [astro-ph.CO].



\bibitem{Starobinsky:1983zz} 
  A.~A.~Starobinsky,
``The Perturbation Spectrum Evolving from a Nonsingular Initially De-Sitter Cosmology and the Microwave Background Anisotropy,''
  Sov.\ Astron.\ Lett.\  {\bf 9}, 302 (1983).
  
\bibitem{Starobinsky:1980te} 
  A.~A.~Starobinsky,
``A New Type of Isotropic Cosmological Models Without Singularity,''
  Phys.\ Lett.\ B {\bf 91}, 99 (1980).
   V.~F.~Mukhanov and G.~V.~Chibisov,
``Quantum Fluctuation and Nonsingular Universe. (In Russian),''
  JETP Lett.\  {\bf 33}, 532 (1981)
  [Pisma Zh.\ Eksp.\ Teor.\ Fiz.\  {\bf 33}, 549 (1981)].

  
\bibitem{Whitt:1984pd} 
  B.~Whitt,
``Fourth Order Gravity as General Relativity Plus Matter,''
  Phys.\ Lett.\ B {\bf 145}, 176 (1984).
  
\bibitem{Linde:2011nh} 
  A.~Linde, M.~Noorbala and A.~Westphal,
``Observational consequences of chaotic inflation with nonminimal coupling to gravity,''
  JCAP {\bf 1103}, 013 (2011)
  [arXiv:1101.2652 [hep-th]].
  
  
\bibitem{Salopek:1988qh}
  D.~S.~Salopek, J.~R.~Bond and J.~M.~Bardeen,
\textit{Designing density fluctuation spectra in inflation},
  Phys.\ Rev.\  \textbf{D40}, 1753 (1989).

  \bibitem{Bezrukov:2008ut}  F.~L.~Bezrukov and M.~Shaposhnikov, ``The Standard
Model Higgs boson as the inflaton,'' Phys.\ Lett.\ B {\bf 659}, 703
(2008) [arXiv:0710.3755 [hep-th]].
 

\bibitem{Okada:2010jf} 
  N.~Okada, M.~U.~Rehman and Q.~Shafi,
``Tensor to Scalar Ratio in Non-Minimal $\phi^4$ Inflation,''
  Phys.\ Rev.\ D {\bf 82}, 043502 (2010)
  [arXiv:1005.5161 [hep-ph]].
  F.~Bezrukov and D.~Gorbunov,
  ``Light inflaton after LHC8 and WMAP9 results,''
  arXiv:1303.4395 [hep-ph].
  
\bibitem{Einhorn:2009bh}
  M.~B.~Einhorn and D.~R.~T.~Jones,
 ``Inflation with Non-minimal Gravitational Couplings in Supergravity,''
  JHEP {\bf 1003}, 026 (2010)
  [arXiv:0912.2718 [hep-ph]].
  
\bibitem{Ferrara:2010yw}
  S.~Ferrara, R.~Kallosh, A.~Linde, A.~Marrani and A.~Van Proeyen,
``Jordan Frame Supergravity and Inflation in NMSSM,'' Phys. Rev. {\bf D82}, 045003
(2010)
  [arXiv:1004.0712 [hep-th]].
  
\bibitem{Lee:2010hj}
  H.~M.~Lee,
  ``Chaotic inflation in Jordan frame supergravity,''
  JCAP {\bf 1008}, 003 (2010)
  [arXiv:1005.2735 [hep-ph]].
  
  \bibitem{Ferrara:2010in} 
  S.~Ferrara, R.~Kallosh, A.~Linde, A.~Marrani, A.~Van Proeyen,
 ``Superconformal Symmetry, NMSSM, and Inflation,''
  Phys.\ Rev.\ D {\bf 83}, 025008 (2011)
  [arXiv:1008.2942 [hep-th]].


\bibitem{Ellis:2013xoa} 
  J.~Ellis, D.~V.~Nanopoulos and K.~A.~Olive,
``A No-Scale Supergravity Realization of the Starobinsky Model,''
  arXiv:1305.1247 [hep-th].

\bibitem{Kallosh:2010ug} 
  R.~Kallosh, A.~Linde,
  ``New models of chaotic inflation in supergravity,''
  JCAP {\bf 1011}, 011 (2010)
  [arXiv:1008.3375 [hep-th]].
  R.~Kallosh, A.~Linde, T.~Rube,
  ``General inflaton potentials in supergravity,''
  Phys.\ Rev.\ D {\bf 83}, 043507 (2011)
  [arXiv:1011.5945 [hep-th]].
  R.~Kallosh, A.~Linde, K.~A.~Olive and T.~Rube,
  ``Chaotic inflation and supersymmetry breaking,''
  Phys.\ Rev.\ D {\bf 84}, 083519 (2011)
  [arXiv:1106.6025 [hep-th]].




\bibitem{Cecotti:1987sa} 
  S.~Cecotti,
``Higher Derivative Supergravity Is Equivalent To Standard Supergravity Coupled To Matter. 1.,''
  Phys.\ Lett.\ B {\bf 190}, 86 (1987).

\bibitem{Ketov:2010qz} 
  S.~V.~Ketov and A.~A.~Starobinsky,
 ``Embedding ($R+R^2$)-Inflation into Supergravity,''
  Phys.\ Rev.\ D {\bf 83}, 063512 (2011)
  [arXiv:1011.0240 [hep-th]].
  S.~V.~Ketov and A.~A.~Starobinsky,
``Inflation and non-minimal scalar-curvature coupling in gravity and supergravity,''
  JCAP {\bf 1208}, 022 (2012)
  [arXiv:1203.0805 [hep-th]].
  S.~V.~Ketov and S.~Tsujikawa,
``Consistency of inflation and preheating in F(R) supergravity,''
  Phys.\ Rev.\ D {\bf 86}, 023529 (2012)
  [arXiv:1205.2918 [hep-th]].
  

  



 
\bibitem{Kallosh:2000ve}
  R.~Kallosh, L.~Kofman, A.~D.~Linde and A.~Van Proeyen,
``Superconformal symmetry, supergravity and cosmology,''
Class.\ Quant.\ Grav.\  {\bf 17}, 4269 (2000)
 [Erratum-ibid.\  {\bf 21}, 5017 (2004)]
 [arXiv:hep-th/0006179].

  \bibitem{delta} R.~Kallosh and A.~Linde,
``Superconformal Generalization of the Chaotic Inflation Model  ${\lambda\over 4} \phi^4 - {\xi\over 2}\phi^2 R$,'' 
  arXiv:1306.3211 [hep-th].
  
\bibitem{Demozzi:2010aj} 
  V.~Demozzi, A.~Linde and V.~Mukhanov,
``Supercurvaton,''
  JCAP {\bf 1104}, 013 (2011)
  [arXiv:1012.0549 [hep-th]].
  
  
  
  
 \end{thebibliography}
\end{document}